# Citations are not opinions: a corpus linguistics approach to understanding how citations are made

Domenic Rosati (dom@scite.ai – scite.ai)

## Abstract

Citation content analysis seeks to understand citations based on the language used during the making of a citation. A key issue in citation content analysis is looking for linguistic structures that characterize distinct classes of citations for the purposes of understanding the intent and function of a citation. Previous works have focused on modeling linguistic features first and drawn conclusions on the language structures unique to each class of citation function based on the performance of a classification task or inter-annotator agreement. In this study, we start with a large sample of a pre-classified citation corpus, 2 million citations from each class of the scite Smart Citation dataset (supporting, disputing, and mentioning citations), and analyze its corpus linguistics in order to reveal the unique and statistically significant language structures belonging to each type of citation. By generating comparison tables for each citation type we present a number of interesting linguistic features that uniquely characterize citation type. What we find is that within citation collocates, there is very low correlation between citation type and sentiment. Additionally, we find that the subjectivity of citation collocates across classes is very low. These findings suggest that the sentiment of collocates is not a predictor of citation function and that due to their low subjectivity, an opinion-expressing mode of understanding citations, implicit in previous citation sentiment analysis literature, is inappropriate. Instead, we suggest that citations can be better understood as claims-making devices where the citation type can be explained by understanding how two claims are being compared. By presenting this approach, we hope to inspire similar corpus linguistic studies on citations that derive a more robust theory of citation from an empirical basis using citation corpora

## 1 Introduction

Citation analysis is an important part of how research is evaluated and research behavior is understood. As has been previously noted (Yousif at al. 2019), citation analysis that doesn't account for how the citation is being made and the content of the citation means that a highly disputed and highly supported paper are counted equally resulting in similar assessments of their impact and other metrics. For this reason, researchers have been motivated to understand the content of citations by characterizing how they are made and finding categories for classifying citations along various dimensions such as their function (i.e. providing background) and polarity (ie. providing criticism) (Alvarez & Gómez, 2016). Researchers attempt to identify language features, such as cue terms or ngrams, that would contribute to a good model for either citation annotation with high inter-annotator agreement or a citation classification with good performance.

 In previous work performance of the modeling of citations was taken to indicate validation of the model and significance of the chosen linguistic features (for instance see Bertin (2016) on making the argument for using ngrams as a feature). This approach can be problematic since the performance of the modeling doesn't necessarily justify the chosen features. The modeling may be capturing hidden features and cues as the primary determinants of a classification. Also, it is not obvious from previous approaches what contribution each feature makes to the resulting classification thus limiting previous works ability to provide an explanation for what makes citation types different.

In this paper, we use this method to discover what language structures contribute uniquely to three citation classes (supporting, disputing, and mentioning citations) from the scite smart citation corpus (see Nicholson et al. 2021). While we present a variety of interesting unique structures to each classification class, we find a result that suggests that previous work on citation content analysis overestimates the power of subjectivity and sentiment when determining whether a citation is supporting or criticizing a cited work. Based on this result we argue that future work on citation content analysis and the theory of citation should focus on citation as a claim-making process rather than previous works which suggest that citations are an opinion-expressing process.

## 2  Methods

For this study we took 6 million citations from the scite Smart Citation dataset as our corpus and selected 2 million citations from each class (supporting, disputing, and mentioning) as subcorpora for comparison. This dataset is over 837 million citation statements and accompanying classifications at the time of writing and represents the largest corpus of citation statements. A corpus of this size is important because it enables citation content analysis at the scale of traditional citation analysis which use similar sized resources such as scopus and web of science (Nicholson et al. 2021). These citation statements (or contexts) contain 1-3 sentences surrounding a citation and are classified into three citation functions based on their rhetorical purpose: to provide support to the cited work, to dispute the cited work, or to simply mention the cited work. The scite classification algorithm uses a deep learning approach where language features are not identified beforehand and the classifications are the result of a model learned from a training set annotated by expert scientists in a variety of fields that has been shown to perform well (Nicholson et al. 2021).

For each class of citation we identified key terms and collocates, in the form of cue terms and ngrams, as they are common language features suggested in citation classification and annotation tasks (Bertin 2016; Kilicoglu 2019; Yousif 2019). For key terms, we used the part of speech tagger available from NLTK and for identifying collocates we used the NLTK bigram and trigram finder with a window size of 5 to accommodate the potentially large citation contexts and selected ngrams with a minimum frequency of 3 occurrences (see Bird et al. 2009). Once key terms and collocates were found across corpora, we created comparison tables that used the log likelihood function, as is common in corpus linguistics significance testing (see Brezina 2018), to calculate the effect of the uniqueness of each key term and collocate per subcorpora. In this context, the log likelihood calculation determines how much more significant a word or phrase is in one corpus versus another based on its likelihood of occurrence.

Once the comparison tables were generated we also performed the following analysis on the collocates in order to contrast language characteristics across classes. First we counted the part-of-speech (POS) tags available in each collocate to understand the distribution of syntactic structures across citation functions. Our hypothesis was that a difference in the distribution of POS tags in each class could tell us about why citations would belong to one class or another. For this we used NLTK's POS tagger which produces tags from the Penn Treebank Tagging project (Bird et al. 2009).

Finally we wanted to test our method on a common hypothesis in citation content analysis that sentiment analysis is an effective means for modeling citation intent (Alvarez & Gómez, 2016). To do this we used NLTK's sentiment analysis feature to calculate the sentiment of each collocate as a way to find out if sentiment within collocates would be a significant linguistic feature in the subcorpora (Bird et al. 2009). While Athar (2014) did show that sentiment is better captured in larger citation contexts we used sentiment analysis on our collocates because this method is commonly used as a key feature in previous citation annotation and classification modeling works (Bertin 2016; Kilicoglu 2019). Finally, in order to understand the role of subjectivity in language features common to citation classes, a necessary precondition for the presence of sentiment (see Liu 2020), we used a pattern-matching based subjectivity measure present in the TextBlob package to understand the subjectivity of the generated collocates (Loria 2020).

|        | Supporting | Disputing | Mentioning |
|--------|------------|-----------|------------|
| Terms | Confirmed, consistent, agrees, supported, shown, demonstrating | Contrast, differ, compared, contradict, failed, differed, inconsistent | Include, used, reviewed, developed, focused, randomized, recommended, implemented |
| Bigrams | "Consistent previous", "Similar reported" | "Contrast previous", "Higher reported", "Lower reported" | "Performed described", "Widely used" |
| Trigrams | "Similar results obtained", "Agreement previous studies" | "Contrast previous reports", "However present study" | "Performed described previously", "Described previously briefly" |

Table 1. Terms and collocates with a log likelihood of belonging to each class above 15.13

|        | Supporting | Disputing | Mentioning |
|--------|------------|-----------|------------|
| Prepositions and subordinating conjunctions (IN) | 1% (Bigrams), 2% (Trigrams) | 3% (Bigrams), 4% (Trigrams) | 1% (Bigrams), 2% (Trigrams) |
| Adjectives (JJ) | 25% (Bigrams), 44% (Trigrams) | 23% (Bigrams), 40% (Trigrams) | 25% (Bigrams), 41% (Trigrams) |
| Adjectives, comparative (JJR) | 0.7% (Bigrams), 2% (Trigrams) | 3% (Bigrams), 8% (Trigrams) | 0.3% (Bigrams), 1% (Trigrams) |
| Adverbs (RB) | 9% (Bigrams), 14% (Trigrams) | 10% (Bigrams), 16% (Trigrams) | 8% (Bigrams), 11% (Trigrams) |
| Adverbs, comparative (RBR) | 0.4% (Bigrams), 0.9% (Trigrams) | 0.9% (Bigrams), 2% (Trigrams) | 0.1% (Bigrams), 0.3% (Trigrams) |

Table 2: Part-of-speech tag distribution in each citation class

## 3  Results

### 3.1  Unique Keywords and Collocates

Table 1 presents examples of the keywords and collocates across citation functions for log likelihoods above the significance of 99.99th percentile indicating terms and phrases highly unique to each subcorpora.

### 3.2  Part-of-speech (POS) Tags

We tabulated the frequency of POS tags and POS sequences for collocates in order to determine if there were unique structures for each citation type. While POS sequences appeared relatively the same across subcorpora, we noted some interesting results where the distribution of POS tags varied quite significantly across sub corpora these are presented in Table 2.

### 3.3  Sentiment and Subjectivity

We wanted to explore the relationship between sentiment and log likelihood of unique phrases. To do this we looked at the Pearson correlation coefficient test whose scores are presented in Table 3.

|          | Supporting | Disputing | Mentioning |
|----------|------------|-----------|------------|
| Bigrams  | r=0.02     | r=-0.007  | r=-0.005   |
| Trigrams | r=0.03     | r=-0.008  | r=-0.006   |

Table 3: Pearson correlation coefficient scores indicating the relationship between collocates in each citation class and sentiment

Which displays the bigrams and trigrams of sentiment versus log likelihood for each citation function. We did the same procedure with the subjectivity score for each collocate and those results are presented in Table 4.

|          | Supporting | Disputing | Mentioning |
|----------|------------|-----------|------------|
| Bigrams  | r=0.01     | r=0.008   | r=-0.001   |
| Trigrams | r=0.02     | r=0.01    | r=-0.009   |

Table 4: Pearson correlation coefficient scores indicating the relationship between collocates in each citation class and subjectivity

## 4  Discussion

Many of the terms and phrases we found would be expected, such as the use of the word 'contrasts', 'contrary', 'differ', and 'contradict' in disputing citations and 'confirmed', 'supported', 'agreement', 'demonstrated' and 'consistent' in supporting citations. These results are generally consistent with the cue terms and ngrams identified in previous literature as contributing uniquely to citation function (Bertin 2016; Murray 2019; Kilicoglu 2019; Teufel 2006).

This indicates that in future work, we could use the methods above to empirically derive cue terms and phrases for explicit annotation and classification tasks. Future world could also look towards a lexical basis for citation classification. However, we cannot rely on the mere presence of these for a complete picture of citation classification since the presence of a term doesn't completely predict the classification of a citation. For example, the collocate "contrast previous studies" which has the highest log likelihood for a trigram unique to the disputing class occurs 537 times in supporting citations, 223 times in mentioning citations and 22,164 times in disputing citations. The presence of lexical terms in other citation classes is common and based on these results, we recommend that future work should not use a lexical basis for characterizing citations classes alone.

Teufel (2006) asserts that "scientific texts are often so full of subjective statements" and incorporates sentiment analysis into much of her and her colleagues' works on citation characterization. In line with this, previous work on citation polarity analysis has been motivated by applying the methods of sentiment analysis to citation classification (Iqbal et al. 2020). However, the most striking aspect of our results is that there is a very low relationship between sentiment, subjectivity, and citation classification on the collocates found most often in each citation class (no correlations are greater than the significant value of 0.05 indicating no statistically significant relationships). This is surprising because intuition would tell us to expect supporting citations to have a positive sentiment and for disputing citations to have a negative citation and for the evaluative claims of citations to have a high subjectivity since they are after all claims about the value of a cited work.

The lack of relationship between subjectivity and citation collocates further corroborates our result since sentiment is often expressed as stated opinion. If citations lack subjectivity features then this suggests that sentiment is not an appropriate means for understanding citation intent or that it is orthogonal to the citation classification task. This contradicts work stating that subjectivity is important to how authors make citations. (Teufel 2006) if citations were understood as opinion-expressing devices, that is how the author feels or what value the author expresses about a cited work, then we should expect to see a much higher subjectivity among collocates. Additionally we should expect to see sentiment as a key feature that differentiates citations function (supporting with positivity, disputing with negativity, and mentioning with neutral) but this is not the case.

Since our results show that this is not the case, what then explains how citations are made? It is instructive to look at Table 2 which shows the POS tag distributions that differentiate citations. While the adverb and adjective forms are found relatively evenly distributed across citation class, comparative adverbs and adjectives are found more often in disputing and supporting classes and much more often in the disputing class. Comparatives indicate that a specific claim is being contrasted. For example: "corroborates earlier studies", a significant collocate for supporting citations compares the current work to "earlier studies" using the time-based comparative "earlier". Another example is the collocate found in disputing citations "study lower [or higher] reported" which provides a quantitative comparison between results. Additionally prepositions and subordinating conjunctions are found more in disputing citations than other classes. These forms indicate the presence of discourse markers setting up a contrastive statement such as "although previous studies" or "'unlike previous 'studies" which are both collocates found significantly in disputing citations.

Based on the above we can see that what characterizes supporting and especially disputing citations is the comparative form which from our results have been judged to have low subjectivity. Comparatives with low subjectivity indicate that a claim (as a statement of fact), is being presented and contrasted with another claim using a specific reference to properties of those claims. This claims-making approach means that a citation intent could be explained by how the comparison between two claims are made. For example, to explain the disputing class for the citation "Some studies reported higher scores for reference in female subjects, [6] but the present results did not support this finding." (Bora & Arabaci, 2009) we can see that the citation in 6 to (Fossati et al., 2003) is treated as a concept symbol for a particular set of results presented in that paper and the disputing class comes from a quantitative comparison. I.e. results in that paper are higher than this paper as implied by "did not support this finding". Looking at the sentiment and subjectivity of this statement as a whole we see that it has a neutral sentiment (0.08 where -1 is negative and 1 is positive) and low subjectivity (0.2 where 0 is very objective and 1 is very subjective). In this view the disputing class is predicted not from an opinion of value expressed but by a comparison of dissimilarity.

The above suggests a claims-making approach is an appropriate way of explaining why citations are made and is more powerful than an opinion-expressing mode of explaining and classifying citations. This explanation also seems in line with our intuitive notions of supporting a claim, comparing a claim by stating it is similar, and disputing a claim, comparing a claim by stating it is different. Future work on citation content analysis should take this into account and a more robust theory of citation based on how claims are compared within citations should be explored. One potential direction is investigating the discourse structure of citations to understand how claims are made specifically and what specific claims are being made as a type of citation reference resolution.

While the preliminary results presented above are interesting, they do not invalidate subjectivity and sentiment as contributors to citation classification as a whole since they are only applied to collocates in this study. Our work does suggest that sentiment in ngrams is not an appropriate feature for citation classification contradicting previous approaches (see Bertin 2016; Kilicoglu 2019). Future work could explore this hypothesis and perform subjectivity and sentiment on entire citation statements rather than just collocates.

Finally, we hope to have shown that a corpus linguistics approach to understanding statistically significant features within citation corpora is a powerful way to analyze citation content at scale and that future studies adopt the methods presented in this paper for identifying language features unique to citations or testing hypotheses about characteristics found in citations.

## Data and Code Availability

The source code for this paper is available at https://github.com/domenicrosati/citation-function-corpus-lingustics. Version 1.0.0, used in this paper, is archived at DOI: 10.5281/zenodo.4608495

Raw citation statement data may be available on request from scite.ai